\documentclass[12pt,preprint]{aastex}

\begin{document}

\title{Ionization Mechanisms of HBLR and Non-HBLR Seyfert 2 Galaxies}
\author{Po-Chieh Yu \& Chorng-Yuan Hwang}
\affil{Graduate Institute of Astronomy, National Central
University, Chung-Li 32001, Taiwan} \email{pcyu@astro.ncu.edu.tw,
hwangcy@astro.ncu.edu.tw}

\begin{abstract}
We investigate the ionization mechanisms for hidden broad-line
region (HBLR) and non-HBLR Seyfert 2 galaxies by comparing some
optical emission line ratios. We note that the [\ion{N}{2}]
$\lambda$6583/H$\alpha$ ratio of the non-HBLR Seyfert 2 galaxies is
significantly higher than that of the HBLR Seyfert 2 galaxies while
other line ratios, such as [\ion{O}{3}]/H$\beta$ and
[\ion{O}{1}]/H$\alpha$ are similar. To probe the origin of this
difference, we explore theoretical results of different ionization
models, such as photoionization, starburst, and shock models. We
find that none of these models can explain the high [\ion{N}{2}]
$\lambda$6583/H$\alpha$ ratio of the non-HBLR Seyfert 2 galaxies
with solar abundance; the high [\ion{N}{2}] $\lambda$6583/H$\alpha$
must be reproduced from enhanced nitrogen abundance. Since nitrogen
overabundance can be achieved from the dredge-up of red supergiants
in the post-main-sequence stage, we suggest that the observed
nitrogen overabundance of the non-HBLR Seyfert 2 might be caused by
stellar evolution, and there could be an evolutionary connection
between the HBLR and non-HBLR Seyfert 2 galaxies.
\end{abstract}


\section{INTRODUCTION}

Seyfert galaxies are classified as radio-quiet active galactic
nuclei (AGNs) and further divided into two subtypes, Seyfert 1 and
Seyfert 2, according to their different optical line widths. Base
on the orientation-based unification model
\citep*{Antonucci85,Antonucci93}, Seyfert 2 galaxies are
considered to be the same objects as Seyfert 1 galaxies but viewed
from a different direction. The detection of polarized broad
permitted emission lines in several Seyfert 2 galaxies further
supported this unification model
\citep*{Tran95,Young19961,Heisler1997,Moran2000}. However,
previous studies showed that only about 40\% - 45\% of Seyfert 2
galaxies have polarized hidden broad line regions (HBLR)
\citep*{Heisler1997, Gu2002}. Spectropolarimetric studies of
Seyfert 2 galaxies showed that the HBLR Seyfert 2 galaxies have
higher luminosities of [\ion{O}{3}], optical, radio and mid
infrared than the non-HBLR Seyfert 2 galaxies
\citep*{Gu2002,Tran2003}. Besides, for several observational
properties, such as $S_{20cm}$/$f_{60}$, $f_{25}$/$f_{60}$, and
$L_{[O III]}$, the HBLR Seyfert 2 galaxies are found to be similar
to Seyfert 1 galaxies, and the non-HBLR Seyfert 2 galaxies are
noted to be more like HII/starburst galaxies
\citep{Gu2002,Deluit}.

It is unclear why some but not all Seyfert 2 galaxies have
detectable HBLRs. Several possibilities have been proposed: (1)
From the ratios of $f_{25}$/$f_{60}$, \citet{Heisler1997}
suggested that the detectability of the HBLR in Seyfert 2 galaxies
is related to the inclination of the torus. (2) Some evolutionary
processes might be at work between the HBLR and non-HBLR Seyfert 2
galaxies \citep*{Tran2003}. (3) \citet{zhang06} found that the
non-HBLR Seyfert 2 galaxies and narrow line Seyfert 1 galaxies
(NLS1s) have similar distribution of black hole masses, accretion
rates and the ratios of $f_{25}$/$f_{60}$. They thus concluded
that the non-HBLR Seyfert 2 galaxies are the counterparts of the
NLS1s at edge-on orientation. (4) The non-HBLR Seyfert 2 galaxies
could be mainly powered by nuclear starbursts rather than
accretion onto the central black hole \citep*{Yu05}. (5) From
X-ray data, \citet{shu07} indicated that the nuclear activity and
obscuration might play an important role in the visibility of
polarized broad lines. (6) \citet{Ho09} showed that the broad line
regions might disappear at low luminosities in advection-dominated
accretion models. (7) \citet{Tran2011} suggested that some Seyfert
2 galaxies could intrinsically lack broad line regions. (8) Some
Seyfert 2 galaxies might be deficient in scattering material.

Emission lines have long been used to distinguish the origin of
active galaxies. For example, optical emission lines of Seyfert 2
galaxies, such as [\ion{O}{3}], [\ion{O}{1}], [\ion{N}{2}], and
[\ion{S}{2}], are usually more stronger than those of
HII/starburst galaxies relative to the recombination lines. For
Seyfert galaxies, photoionization from a power-law continuum,
presumedly from the central AGN, is proposed to be responsible for
the ionization mechanisms and is generally successful in
reproducing observed optical lines \citep*{veilleux1987}. If the
non-HBLR Seyfert 2 galaxies were powered by nuclear starburst
activities, one important question is how the narrow line regions
(NLRs) of the non-HBLR Seyfert 2 galaxies are ionized. However, we
note that \citet{Terlevich1992} have showed that most of line
ratios of the Seyfert 2 galaxies could be reproduced by starburst
models. Besides, a photoionization model with power-law continuum
might fail to reproduce all observed properties in some cases. For
example, observations of [\ion{O}{3}] suggested that the electron
temperature could be up to $\sim$22,000 K \citep{Tadhunter1989}
while the photoionization model could only give a lower one
($\sim$11,000 K). Therefore, some other physical processes, such
as shock-heating and starburst, might also play a significant role
even in a photoionization dominated model \citep{Kraemer1998}.
These results suggest that we need to consider different
ionization mechanisms when we investigate the origin of the
ionization for Seyfert 2 galaxies.

In this paper, we investigate and compare possible ionization
mechanisms of the NLRs for the HBLR and non-HBLR Seyfert 2
galaxies by considering photoionization, shock-wave heating, and
starburst models. We calculate line ratios of the photoionization
model with the Cloudy program of version 08.00, last described by
\citet{Ferland1998}. The shock models are obtained from the
library calculated by \citet{Allen2008} with the MAPPING III code,
and the starburst models are obtained from \citet{Kewley2001_1}.

\section{LINE RATIOS OF HBLR AND NON-HBLR SEYFERT 2 GALAXIES}

\subsection{Diagnostic Diagrams}
Active galaxies, such as starburst and Seyfert 2, show narrow
emission lines in their spectra. The line ratio [\ion{O}{3}]
$\lambda$5007/H$\beta$ $\geq$ 3 was found to be a good criterion
to separate Seyfert 2 and starburst or \ion{H}{2}-like galaxies
\citep*{Shunder1981}. However, some starburst galaxies were also
found to have [\ion{O}{3}] $\lambda 5007/H\beta \geq 3$
\citep*{Osterbrock1985}. Other line ratios, such as
[\ion{N}{2}]~$\lambda$6583/H$\alpha$,
[\ion{O}{1}]~$\lambda$6300/H$\alpha$ and
[\ion{S}{2}]~$\lambda$$\lambda$6716,~6731/H$\alpha$, are also
needed to separate the starburst and Seyfert 2 galaxies
\citep*{BPT81, veilleux1987}. In Figure~1, we show the diagnostic
diagram of these line ratios; the data of the HBLR and non-HBLR
Seyfert 2 galaxies are collected from literatures and listed in
Table~1. The data of \ion{H}{2}-like galaxies and LINERs are
adopted from \citet{Ho1997}.

Beside the well-known separation between the starburst and Seyfert
2 galaxies, we note that the distribution of the non-HBLR Seyfert
2 is very different from that of the HBLR ones. The difference is
mainly caused by the different line ratio of [\ion{N}{2}]
$\lambda$6583/H$\alpha$. Both Kolmogorov-Smirnov test ($P=99.7\%$)
and Student's T-test ($P=99.9\%$) show that the [\ion{N}{2}]
$\lambda$6583/H$\alpha$ distribution of the HBLR and non-HBLR
Seyfert 2 galaxies are significantly different. This indicates
that the HBLR and non-HBLR Seyfert 2 galaxies have different
physical conditions. In order to understand the physical
properties of these Seyfert 2 galaxies, we compare the observed
data with different ionization mechanisms in the following.

\subsection{Models}
Seyfert galaxies are usually considered to be photoionized by
central AGNs. The ionization source is suggested to be a power-law
continuum \citep*{Koski1978}. In order to compare with the
observed data, we use the Cloudy program to produce the line
ratios. The Cloudy package is designed to simulate photoionization
of clouds with different incident continuum. We assume that the
incident ionizing continuum is a power law with an index $p=1.4$
and the distance from the cloud to the central continuum is set to
be 100 pc. The derived ionization parameter $\Gamma$ varies from
10$^{-1.5}$ to 10$^{-3.5}$ and the hydrogen density $n_{H}$ varies
from 10$^{2}$ to 10$^{5}$~cm$^{-3}$.

Figure~2 shows the results of Cloudy models with different
nitrogen abundances. Figure~2a shows that the relative emission
between [\ion{O}{3}] and [\ion{O}{1}] is not very sensitive to the
nitrogen abundance as expected. Figure~2b indicates that the high
[\ion{N}{2}] $\lambda$6583/H$\alpha$ ratio of the non-HBLR Seyfert
2 galaxies can be reproduced by increasing the nitrogen abundance
to five times solar abundance (5N$_{\sun}$). We note that if we
simply increase all metal abundance, the [\ion{O}{3}] emission
would also increase as well and would not reproduce the observed
[\ion{N}{2}]/[\ion{O}{3}] ratio. This result suggests that the
non-HBLR Seyfert 2 galaxies might have higher N/O relative
abundance than the HBLR Seyfert 2 galaxies. However, as shown in
Figure~2c, the [\ion{S}{2}] emission can not be reproduced with
both (N$_{\sun}$) and 5N$_{\sun}$ abundance.

To investigate the origin of the [\ion{S}{2}] emission, we compare
the emission of [\ion{N}{2}] with that of [\ion{S}{2}] in
Figure~3. There is a significant correlation between [\ion{N}{2}]
$\lambda$6583/H$\alpha$ and [\ion{S}{2}]$\lambda$$\lambda$6716,
6731/H$\alpha$ for both HBLR and non-HBLR seyfert 2 galaxies as
shown in Figure~3. This indicates that the emission of both
[\ion{N}{2}] and [\ion{S}{2}] must have the same origin. We also
note that the correlations might have different slopes for both
kinds of galaxies; the non-HBLR Seyfert 2 galaxies seems to have a
slightly steeper slope than that of the HBLR Seyfert 2 galaxies.
Furthermore, the photoionization model is difficult to explain the
high [\ion{N}{2}]. This suggests that some other mechanisms might
be operative for producing the observed [\ion{S}{2}] and
[\ion{N}{2}] emission.

One other possible excitation mechanism for the NLRs of Seyfert 2
galaxies is shock-wave heating. \citet{Allen2008} has presented a
library of radiative shock models calculated using the MAPPINGS
III code. The parameters in the library have broad ranges with the
pre-shock density $n$ from 0.01 to 1000~cm$^{-3}$, the shock
velocity $v$ from 100 to 1000 km~s$^{-1}$, the pre-shock magnetic
fields $B$ from 10$^{-10}$ to 10$^{-4}$~G, and the magnetic
parameter $B$/$n$$^{1/2}$ from 10$^{-4}$ to 100~$\mu$G~cm$^{3/2}$.
The library of MAPPINGS III code is available in the
internet.\footnote[1]{http://www.ifa.hawaii.edu/$\sim$kewley/Mappings/index.html}
We compare the observed line ratios with the results from the
library.


The shock model with the standard solar abundance can reproduce
the distribution of the [\ion{O}{1}] and [\ion{S}{2}] emission. On
the other hand, the [\ion{N}{2}] emission of the non-HBLR Seyfert
2 galaxies can be reproduced only by increasing metal content to
two times solar abundance in the shock models. This result further
suggests that the nitrogen abundance of the non-HBLR Seyfert 2
galaxies are higher than that of the HBLR ones. However, it is
noted that the shock models can not produce enough [\ion{O}{3}]
emission. By adding ionized pre-shock gas in the shock models
\citep{Allen2008} might be able to produce high enough
[\ion{O}{3}] emission but would fail to produce the observed
[\ion{N}{2}] emission. These results show that shock models are
unable to reproduce these observed line ratios simultaneously.

Another possible ionization mechanism is starburst. A large
library of starburst models has been presented by
\citet{Kewley2001_1}. However, as shown in Figure~1, only three
sources lie on the extreme starburst region. It is obvious that
the [\ion{N}{2}] and [\ion{O}{3}] line emission of the non-HBLR
Seyfert 2 galaxies can not be reproduced by the starburst models.
Furthermore, if the [\ion{N}{2}] emission is from starburst
region, high [\ion{O}{2}] emission would be expected. However, the
averaged [\ion{O}{2}]/[\ion{O}{3}] ratio is $0.23\pm0.16$ and
$0.26\pm0.09$ for the HBLR and non-HBLR Seyfert 2 galaxies
\citep{Gu2006}, indicating the contribution from star formation is
similar for the HBLR and non-HBLR Seyfert 2 galaxies. Besides,
\citet{Kewley2001_1} indicated that the contribution from
supernovae to the starburst models is $\ll$ ~20\% and can be
neglected. Therefore, the starburst models would hardly explain
these line ratios even including supernova contribution.

Based on these results, all the line ratios can not be reproduced
by one single model. This might not be unreasonable since NLRs
could be composed of clouds with different ionization conditions.
For example, the [\ion{O}{3}] emission could come from highly
ionized regions that powered by AGNs while [\ion{N}{2}] emission
come from lower ionized regions. One possibility to explain the
observed line ratios is a composite model of an AGN continuum and
starburst and/or shock. However, the [\ion{N}{2}]/H$\alpha$ line
ratios produced by these three single models with solar abundance
are all much less than observations, so any combination of these
three models can not produce enough [\ion{N}{2}] emission for the
non-HBLR Seyfert 2 galaxies. This result suggests that the high
[\ion{N}{2}]/H$\alpha$ ratios must be due to over abundance of
nitrogen in the non-HBLR Seyfert 2 galaxies.

\section{DISCUSSION}
It is not unusual for Seyfert 2 galaxies to show enhanced nitrogen
abundance. Previous studies showed that the nitrogen abundance of
Seyfert 2 galaxies can reach to 3.5--5 solar abundance
\citep*{Storchi1990}. Our results further show that the non-HBLR
Seyfert 2 galaxies have higher nitrogen abundance than  the HBLR
Seyfert 2 galaxies. Comparing these results, we suggest that the
high nitrogen abundance of the Seyfert 2 galaxies in early studies
might be mainly caused by the non-HBLR Seyfert 2 galaxies.

The difference between the HBLR and non-HBLR might be caused by
their stellar evolution. Based on evolutionary models of starburst
activities, \citet{Matteucci1993} showed that the N/O relative
abundance reaches a maximum value at about 3 $\times$ 10$^{8}$
years. The nitrogen could be dredged up from red supergiants in the
post-main-sequence stage, and this would result in the nitrogen
overabundance. Therefore, the super-solar nitrogen abundance of the
non-HBLR Seyfert 2 galaxies may imply that the starburst activities
in the non-HBLR Seyfer 2 galaxies are in the post-main-sequence
stage.

The high abundance of the non-HBLR Seyfert 2 galaxies implies that
there could be an evolutionary connection between the non-HBLR and
HBLR Seyfert 2 galaxies. AGNs have been suggested to be related to
circumnuclear star formation. Accreted gas is transported to the
central regions to fuel the AGNs; the transported gas accumulates
around the nucleus and might also trigger star formation. The
super-solar abundance suggests that the non-HBLR Seyfert 2
galaxies are in older stages of stellar evolution than the HBLR
Seyfert 2 galaxies. This would also imply that the gas around the
nucleus of the non-HBLR Seyfert 2 galaxies might have diminished,
which would cause low accretion rates for the non-HBLR Seyfert 2
galaxies. This is consistent with the results found by
\citet{Gu07}, which showed that the accretion rates of non-HBLR
Seyfert 2 galaxies are lower than those of HBLR Seyfert 2
galaxies. When the accretion rate is below some thresholds, the
broad line regions could disappear \citep{Nicastro03}; therefore,
no polarized broad lines are observed in the non-HBLR Seyfert 2
galaxies.

\section{SUMMARY}
We investigate the ionization mechanisms of the HBLR and non-HBLR
Seyfert 2 galaxies by comparing the photoionization , shock, and
starburst results with observations. A possible explanation for
all observed emission might require combination of different
models; for example, the [\ion{O}{3}] emission could come from
highly ionized regions that powered by AGNs while [\ion{N}{2}]
emission come from lower ionized regions by starburst or shock
waves. However, our results show that the non-HBLR Seyfert 2
galaxies must have higher N/O relative abundance than the HBLR
Seyfert 2 galaxies. The high N/O relative abundance could
originate from stellar evolution and imply an evolutionary
connection between the HBLR and non-HBLR Seyfert 2 galaxies. As a
HBLR Seyfert 2 galaxy evolves, the gas around the nucleus become
diminished, resulting in low accretion rates and depleting the
broad line regions; this evolution might transfer HBLR Seyfert 2
galaxies to non-HBLR ones.

\acknowledgments We thank the referee for constructive comments.
We are also grateful to the groups and persons who provide the
Cloudy code and the MAPPINGS III library. This work was partially
supported by the National Science Council of Taiwan (grants NSC
99-2112-M-008-014-MY3 and NSC 99-2119-M-008-017). This research
has made use of the NASA/IPAC Extragalactic Database (NED) which
is operated by the Jet Propulsion Laboratory, California Institute
of Technology, under contract with the National Aeronautics and
Space Administration.

\clearpage
\begin{deluxetable}{lrrrrr}\tabletypesize{\small}
\tablecolumns{6} \tablewidth{0pt}\tablecaption{Line Ratios of HBLR
and Non-HBLR Seyfert 2 Galaxies} \tablehead{ \colhead{Name} &
\colhead{[\ion{O}{3}] $\lambda$5007/H$\beta$}   &
\colhead{[\ion{O}{1}] $\lambda$6300/H$\alpha$} &
\colhead{[\ion{N}{2}] $\lambda$6583/H$\alpha$} &
\colhead{[\ion{S}{2}] $\lambda$6720/H$\alpha$} & \colhead{Ref.}}
  \startdata
  \cutinhead{HBLR Seyfert 2 Galaxies}
IC 3639  &0.98   &-1.10   &-0.11   &-0.31  &(1)\\
IC 5063  &0.99   &-0.93   &-0.20   &-0.28  &(1)\\
IRAS 01475-0740 &0.72    &\nodata    &-0.31  &\nodata  &(2) \\
IRAS 02581-1136 &1.19    &\nodata    &-0.09  &\nodata  &(2) \\
IRAS 04385-0828 &0.33    &\nodata    &-0.19  &\nodata  &(2) \\
IRAS 05189-2524 &1.53    &-1.12   &0.03  &-0.72 &(3) \\
IRAS 11058-1131 &0.96    &-1.28   &-0.42   &-0.59 &(4)\\
IRAS 15480-0344 &1.28    &-1.30   &-0.18   &-0.66 &(4)\\
IRAS 18325-5926 &0.66    &\nodata      &-0.20   &\nodata    &(1) \\
IRAS 20460+1925 &0.78    &-1.23   &-0.28   &-0.62 &(5) \\
IRAS 22017+0319 &0.97    &\nodata    &-0.32  &\nodata  &(2)\\
MCG 5-23-16 &-0.40   &-0.28   &-0.11   &-0.07 &(6) \\
Mrk 1210 &1.02    &-0.66   &-0.34   &-0.73 &(7) \\
Mrk 3    &1.17    &\nodata  &-0.01  &\nodata  &(2)\\
Mrk 348  &1.14    &\nodata  &-0.15  &\nodata  &(2)\\
Mrk 573  &1.01    &-0.96   &-0.11   &-0.29 &(8) \\
NGC 1068 &1.11    &-1.06   &-0.12   &-0.62 &(9) \\
NGC 2273 &0.76    &-0.92   &-0.07   &-0.33 &(9) \\
NGC 424  &0.66    &-1.20   &-0.41   &-0.81 &(10) \\
NGC 4388 &1.05    &-0.80   &-0.24   &-0.21 &(9) \\
NGC 4507 &0.88    &-0.85   &-0.24   &-0.42 &(1) \\
NGC 5252 &0.84    &-0.37   &-0.06   &-0.09 &(8) \\
NGC 5347 &0.95    &\nodata &-0.11   &\nodata &(2) \\
NGC 591  &0.99    &-0.87   &0.05    &-0.22 &(6)\\
NGC 5929 &0.57    &-0.58   &-0.22   &-0.16 &(8)\\
NGC 5995 &1.15    &-0.72   &0.15    &\nodata &(1)\\
NGC 7212 &1.07    &-0.78   &-0.14   &-0.34 &(1)\\
NGC 7674 &1.03    &-1.30   &-0.01   &-0.49 &(1)\\
NGC 7682 &0.97    &-0.58   &0.01    &-0.15 &(8)\\
NGC 788  &1.30    &-0.37   &-0.10   &-0.19 &(10)\\
IRAS 00521-7054   &0.99    &-0.88   &-0.03 &-0.46 &(11)\\
IRAS 23060+0505   &0.97    &\nodata &-0.27 &\nodata &(2)\\
NGC 5506 &0.88    &-0.84   &-0.09   &-0.13 &(1)\\

\cutinhead{Non-HBLR Seyfert 2 Galaxies}
IRAS 00198-7926 &0.50  &-1.25     &-0.30  &-0.48   &(1)\\
IRAS 03362-1642 &0.79  &\nodata   &-0.20  &\nodata &(2)\\
IRAS 04229-2528 &0.60  &\nodata   &-0.01  &\nodata &(2)\\
IRAS 04259-0440 &0.30  &-0.85     &-0.11  &-0.44   &(1)\\
IRAS 08277-0242 &1.15  &\nodata   &0.11   &\nodata &(2)\\
IRAS 13452-4155 &0.92  &-0.78     &-0.21  &-0.37   &(11)\\
IRAS 20210+1121 &0.79  &-0.99     &-0.17  &-0.50   &(12)\\
IRAS 23128-5919 &0.48  &-10.00    &-0.49  &-0.70   &(1)\\
Mrk 1066        &0.59  &-1.03     &-0.06  &-0.41   &(6)\\
Mrk 334         &0.23  &-1.28     &-0.23  &-0.55   &(8)\\
NGC 1144        &0.88  &-0.63     &0.27   &-0.08   &(8)\\
NGC 1241        &0.74  &-0.14     &-0.03  &-0.27   &(10)\\
NGC 1320        &1.05  &-0.90     &-0.15  &-0.38   &(10)\\
NGC 1358        &1.05  &-0.59     &0.30   &-0.02   &(9)\\
NGC 1386        &1.57  &-1.10     &0.20   &-0.14   &(1)\\
NGC 1667        &0.88  &-0.62     &0.38   &-0.02   &(9)\\
NGC 1685        &0.87  &\nodata   &-0.08  &\nodata &(2)\\
NGC 3079        &0.62  &-0.74     &0.20   &-0.07   &(9)\\
NGC 3281        &1.00  &-1.03     &-0.01  &-0.26   &(1)\\
NGC 3362        &0.92  &-0.65     &0.38   &-0.06   &(8)\\
NGC 34          &0.50  &-0.93     &0.10   &-0.33   &(3)\\
NGC 3982        &1.33  &-0.49     &-0.06  &-0.24   &(9)\\
NGC 4501        &0.73  &-0.72     &0.32   &-0.03   &(9)\\
NGC 5135        &0.69  &-1.19     &-0.04  &-0.45   &(1)\\
NGC 5194        &0.95  &-0.80     &0.46   &-0.07   &(9)\\
NGC 5256        &0.62  &-1.25     &-0.25  &-0.44   &(3)\\
NGC 5283        &0.79  &-0.58     &-0.06  &-0.11   &(8)\\
NGC 5643        &1.11  &-0.73     &0.06   &-0.15   &(1)\\
NGC 5695        &1.02  &-0.55     &0.17   &0.03    &(6)\\
NGC 6300        &0.96  &-0.52     &0.35   &-0.03   &(10)\\
NGC 7172        &1.00  &-0.78     &0.00   &0.01    &(10)\\
NGC 7582        &-0.20 &-1.68     &-0.50  &-0.62   &(1)\\
IRAS 19254-7245 &0.72  &-0.86     &0.06   &-0.14   &(1)\\

\enddata
\tablecomments{Column 2 - Column 5: Line ratio presented in
logarithm. Column 6: Reference number - (1)\citet{Kewley2001};
(2)\citet{Grijp1992}; (3)\citet{veilleux1995};
(4)\citet{Osterbrock1985}; (5)\citet{Frogel1989};
(6)\citet{veilleux1987}; (7)\citet{Terlevich1991};
(8)\citet{Osterbrock1993}; (9)\citet{Ho1997};
(10)\citet{vaceli1997}; (11)\citet{vader1993};
(12)\citet{perez1989}.}
\end{deluxetable}

\clearpage
\begin{figure}
\epsscale{1} \figurenum{1}\plotone{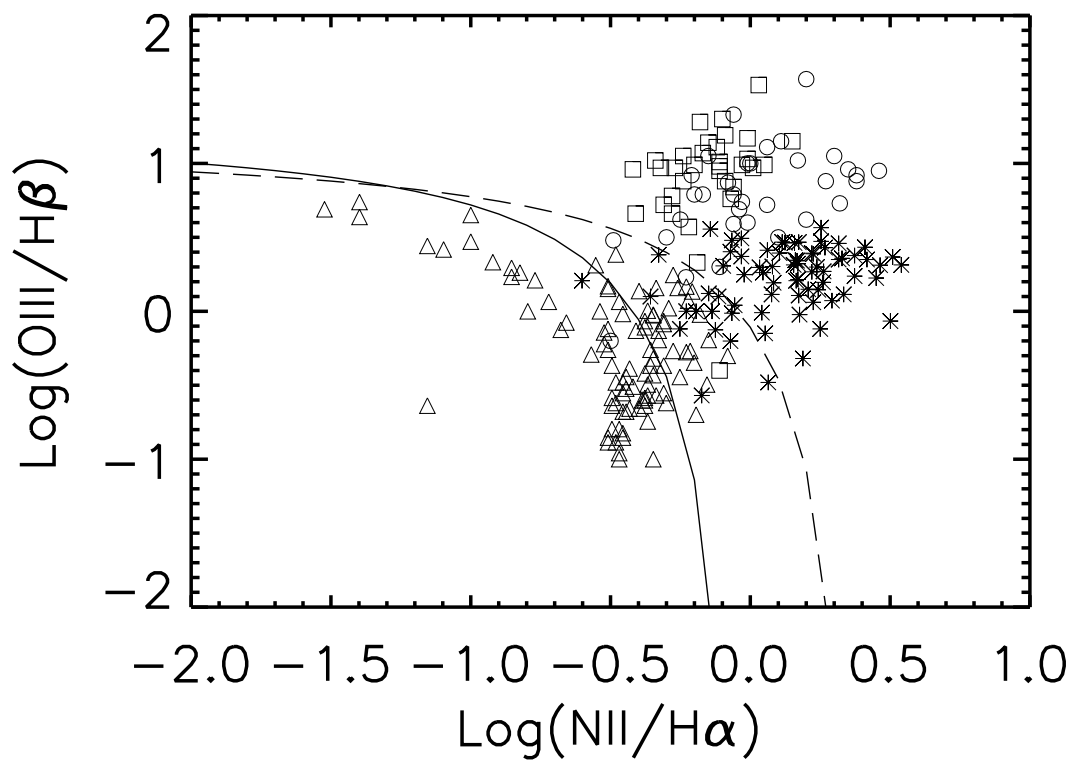} \caption{Diagnostic
diagrams of [\ion{N}{2}] and [\ion{O}{3}] line ratios. The HBLR
Seyfert 2 galaxies are shown as squares, the non-HBLR Seyfert 2 as
open circles, HII like as triangles and LINER as asterisks. The
solid curve represents the dividing line of star forming galaxies
from Seyfert-HII composite sources \citep{Ka03}; and the dashed
line represents the boundary of extreme starburst with super-solar
abundance \citep{Kewley2006}.}
\end{figure}

\clearpage
\begin{figure}
\epsscale{0.6} \figurenum{2}\plotone{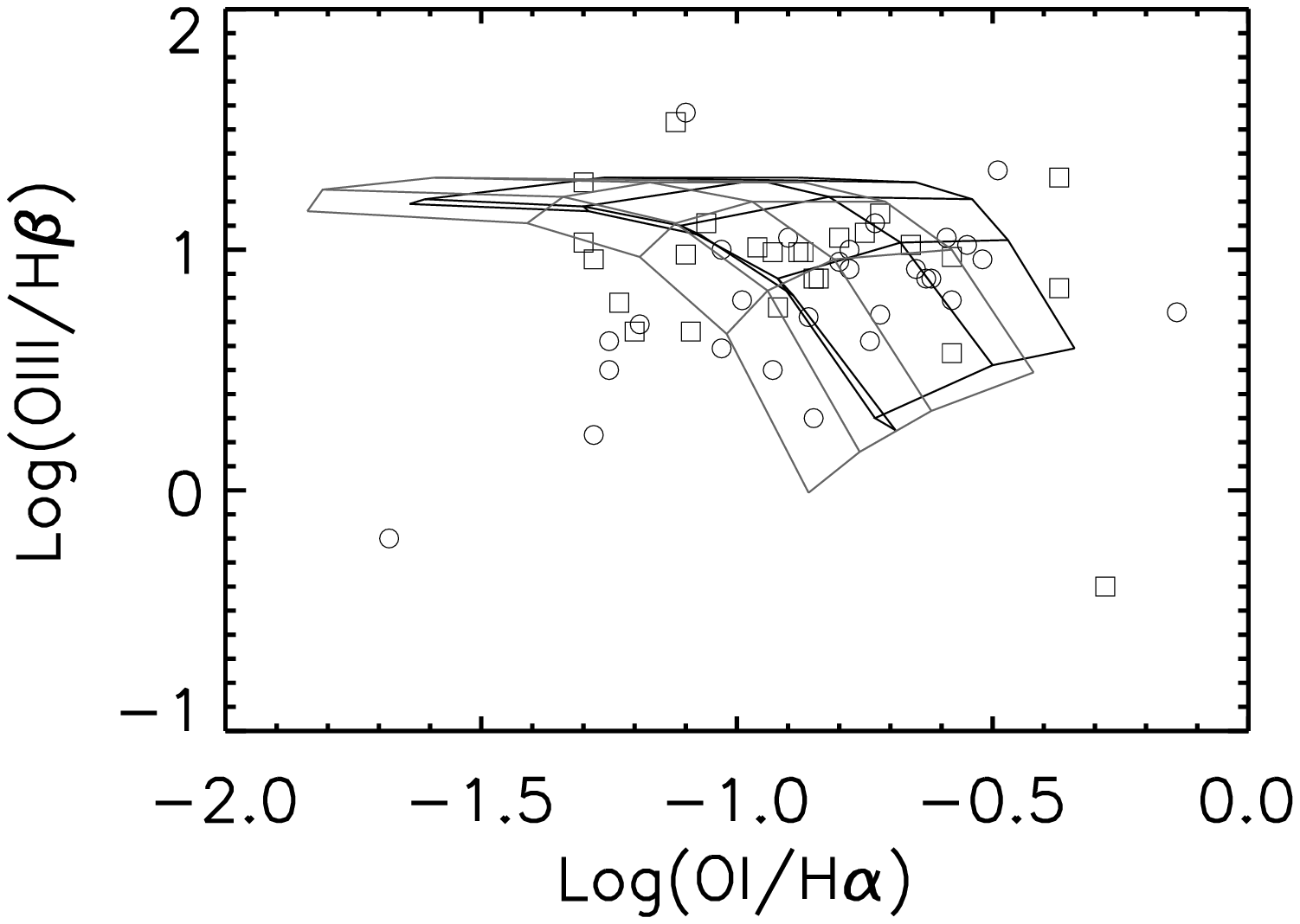}
\figurenum{2}\plotone{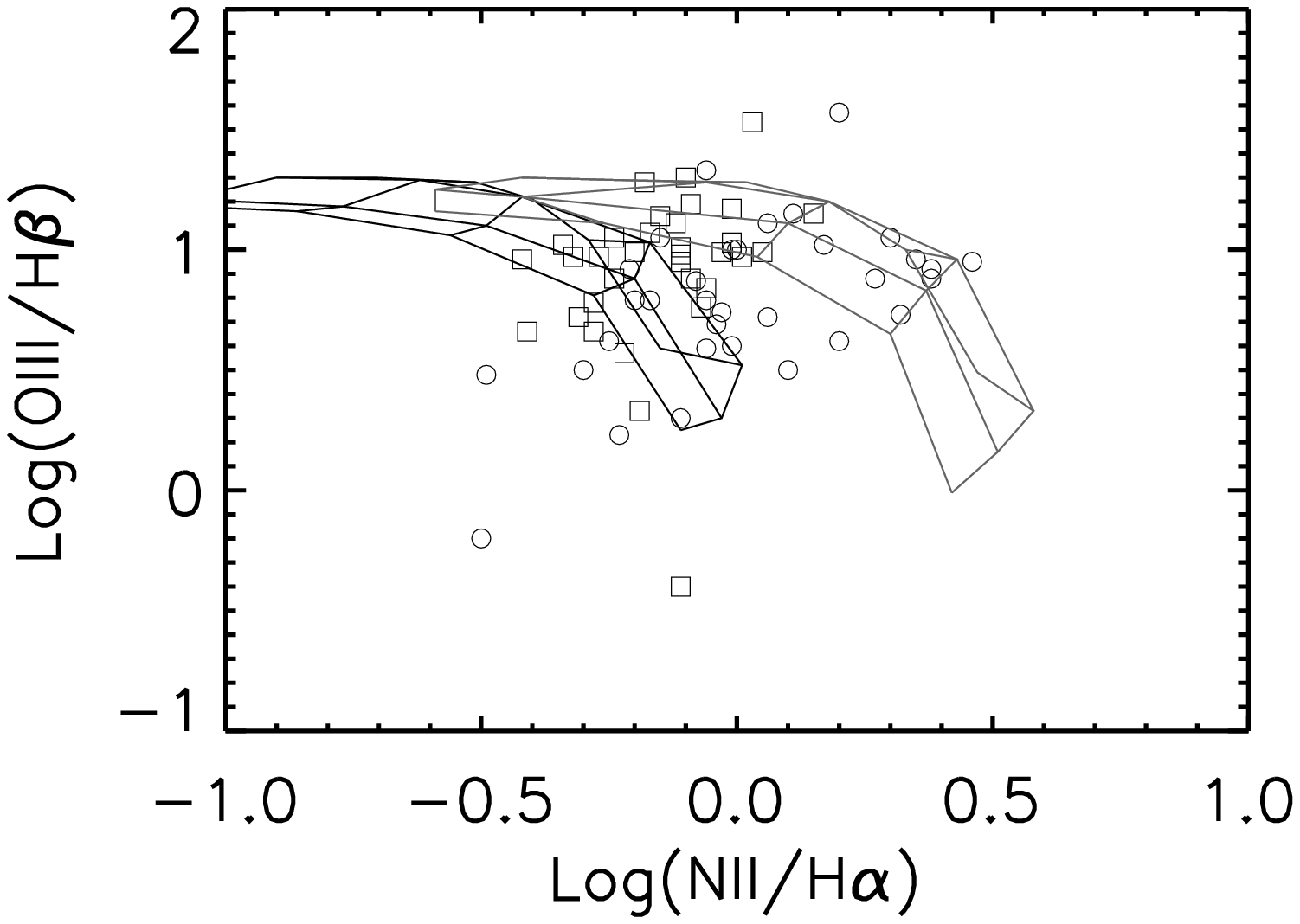}
\figurenum{2}\plotone{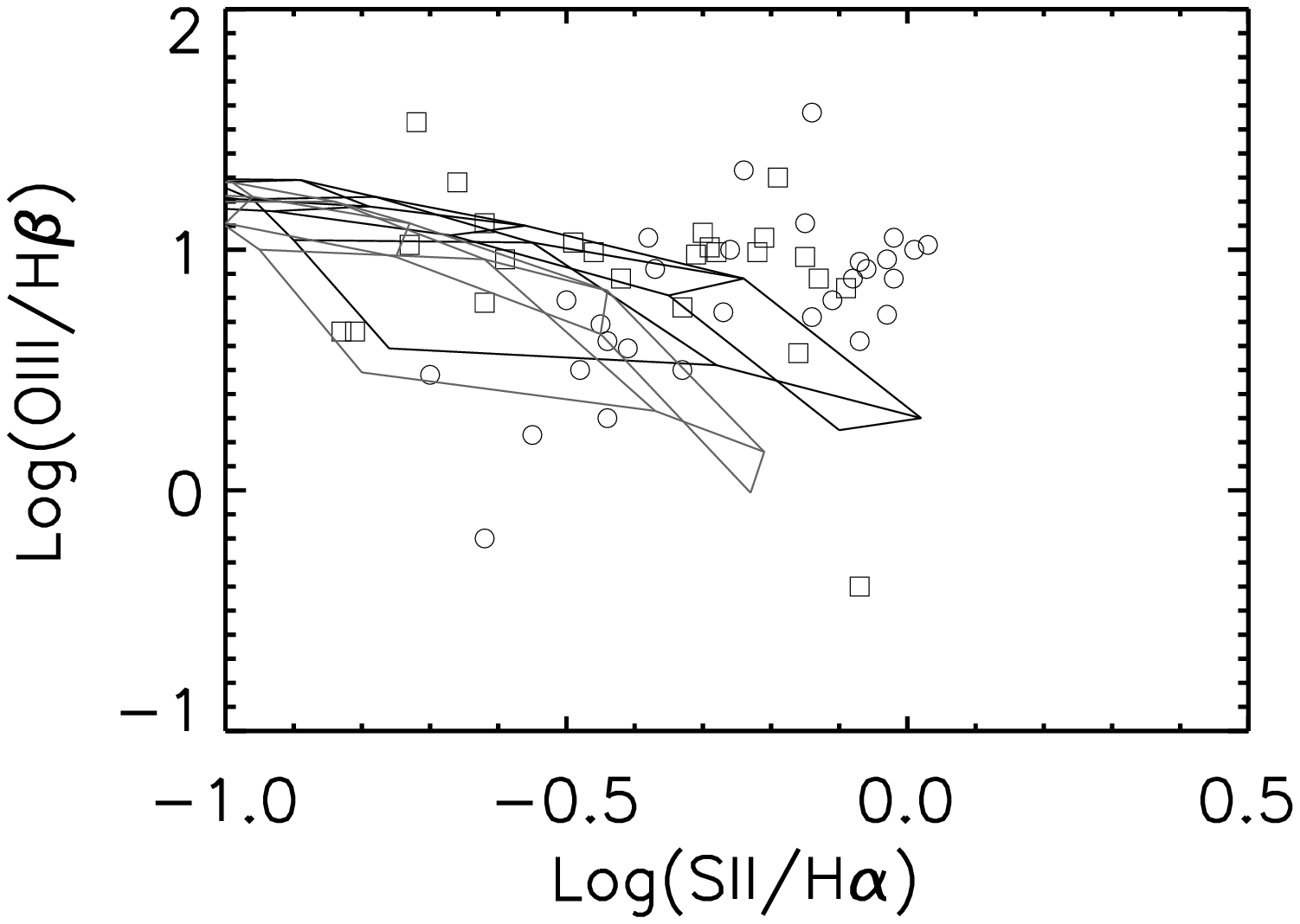}\caption{Results of photoionization
models. Symbols are the same as those of Figure~1. The results
with the standard solar abundance of nitrogen (N$_{\sun}$) are
shown in black curves, and those with five times solar abundance
of nitrogen (5N$_{\sun}$) are shown in grey. The ionization
parameter $\Gamma$ varies from 10$^{-1.5}$ to 10$^{-3.5}$
(upper-left to lower-right for each model), and the density
n$_{H}$ varies from 10$^{2}$ to 10$^{5}$ (left to right for each
model).}
\end{figure}

\clearpage
\begin{figure}
\epsscale{1} \figurenum{3}\plotone{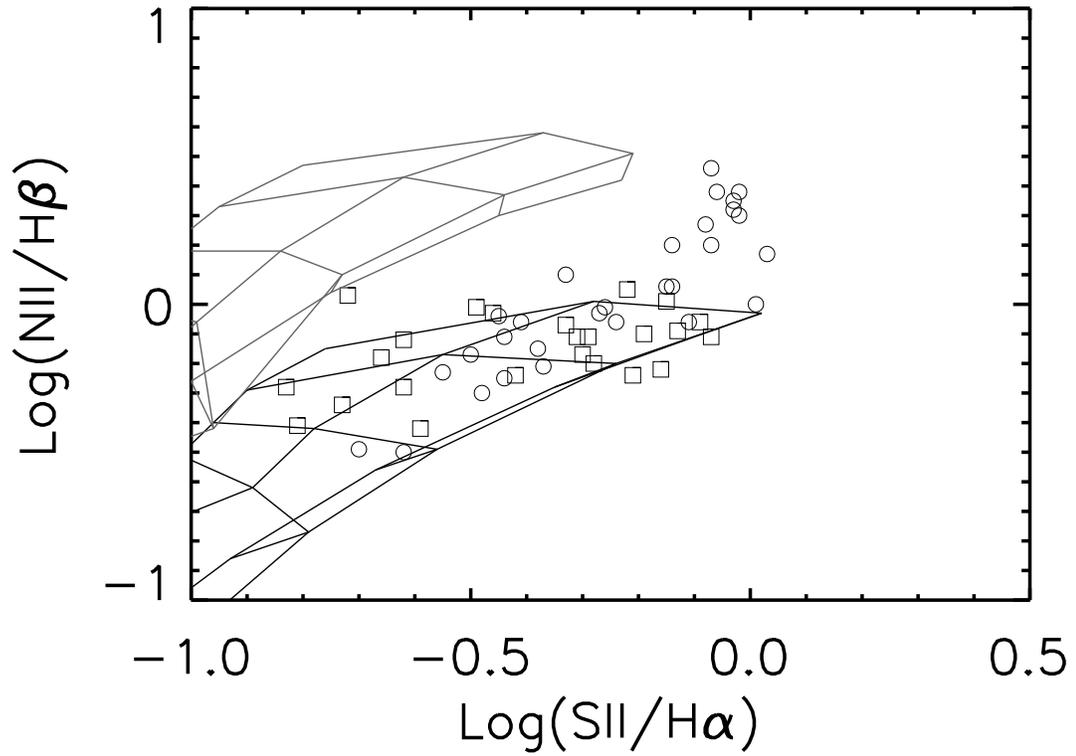} \caption{Correlation
between the [\ion{N}{2}] and [\ion{S}{2}] emission. The results of
photoionization models are also plotted in the diagram for
comparison. Symbols are the same as those of Figure~1. The
ionization parameter $\Gamma$ varies from 10$^{-1.5}$ to
10$^{-3.5}$ (lower-left to upper-right for each model), and the
density n$_{H}$ varies from 10$^{2}$ to 10$^{5}$ (left to right
for each model).}
\end{figure}

\end{document}